# Landau Modes are Eigenmodes of Stellar Systems in the Limit of Zero Collisions

C. S. Ng

*University of Alaska Fairbanks, Fairbanks, Alaska 99775, USA*

ORCID: 0000-0003-1861-5356

A. Bhattacharjee

*Department of Astrophysical Sciences and Princeton Plasma Physics Laboratory, Princeton University, Princeton, NJ 08540, USA*

ORCID: 0000-0001-6411-0178

**Abstract**

We consider the spectrum of eigenmodes in a stellar system dominated by gravitational forces in the limit of zero collisions. We show analytically and numerically using the Lenard-Bernstein collision operator that the Landau modes, which are not true eigenmodes in a strictly collisionless system (except for the Jeans unstable mode), become part of the true eigenmode spectrum in the limit of zero collisions. Under these conditions, the continuous spectrum of true eigenmodes in the collisionless system, also known as the Case-van Kampen modes, is eliminated. Furthermore, since the background distribution function in a weakly collisional system can exhibit significant deviations from a Maxwellian distribution function over long times, we show that the spectrum of Landau modes can change drastically even in the presence of slight deviations from a Maxwellian, primarily through the appearance of weakly damped modes that may be otherwise heavily damped for a Maxwellian distribution. Our results provide important insights for developing statistical theories to describe thermal fluctuations in a stellar system, which are currently a subject of great interest for N-body simulations as well as observations of gravitational systems.

*Unified Astronomy Thesaurus concepts*: Galaxy kinematics (602); Galaxy dynamics (591); Galaxy physics (612);

**1 Introduction**

Stars in galaxies are believed to be collisionless. Textbook estimates indicate that for a typical number of stars $N \sim 10^{11}$ in a galaxy, the typical mean free path is of the order of $10^{14}$ parsecs and the time between collisions is of the order of $10^{18}$ years, which exceeds the life of the Universe by about eight orders of magnitude (Binney & Tremaine 2008). Hence, it appears safe to ignore the effects of collisions over the lifetime of a galaxy. It is recognized, of course, that this conclusion on the absence of collisions is not always valid and that collisions can be significant in specific contexts (Saslaw 1985, Binney & Tremaine 2008). Still, by and large, the collisionless Boltzmann equation (also known as the Vlasov equation) has remained the standard point of departure for studies of stellar dynamics.

Collisionless stellar systems governed by Newton's law of gravitation and electrostatic plasma systems governed by Coulomb's law have an essential property in common: they are both governed by long-range inverse-square forces. This similarity has been a source of useful cross-fertilization between the fields of plasma physics and stellar dynamics, which have shared many mathematical concepts and physical models, such as Case-van Kampen modes and Landau damping.

The single-particle probability distribution function (DF) $F(\mathbf{x}, \mathbf{v}, t)$ is a function of position $\mathbf{x}$, velocity $\mathbf{v}$, and time $t$ and obeys the time-dependent Vlasov equation. $F(\mathbf{x}, \mathbf{v}, t)$ is assumed to be a smooth function in six-dimensional phase space. This phase-space probability density smooths over particle discreteness and obeys Liouville's theorem, a distinguishing property of all Hamiltonian systems. However, if one starts with a smooth blob in phase space and solves the self-consistent Vlasov-Poisson equations, the blob quickly deforms to exhibit whorls and tendrils (like cream in coffee) but always in a manner that is consistent with Liouville's theorem. Thus, the DF generates strong filamentary structures in phase space even if the initial conditions are smooth. In N-body computer simulations, these filamentary structures often display such strong gradients in velocity space that it becomes difficult to resolve them. Higher-order derivatives of $F(\mathbf{x}, \mathbf{v}, t)$ become increasingly large. No matter how small the collisions are in the physical system, there comes a time in initial-value studies when it is no longer possible to neglect collisional effects that depend on higher-order derivatives of $F(\mathbf{x}, \mathbf{v}, t)$ as they do, for example, in the Fokker-Planck collision operator, which is derived under the assumption that collisional processes are Markovian (Chandrasekhar 1943, Rosenbluth, MacDonald, & Judd 1957, Kandrup 1983, Saslaw 1985, Binney & Tremaine 2008, Chavanis 2008).

Recently, Lau & Binney (2021) have considered the problem of thermal fluctuations of star clusters and galaxies around statistical equilibria. They advocate the use of the Case-van Kampen modes (van Kampen 1955, Case 1959) as normal modes in such a theory, instead of the Landau modes (Landau 1946) used in previous studies by Kalnajs (1965), Toomre (1964, 1981) and Weinberg (1993, 1994, 1998). The purpose of this paper is to revisit the perspectives of Lau and Binney in light of some recent results in plasma physics regarding the effect of weak collisions on the eigenmode spectrum that do not yet appear to have found applications in the stellar dynamics literature. Specifically, we demonstrate that if one views the collisionless system as the zero-collision limit of the collisional system, the Case-van Kampen modes are eliminated, and the Landau modes become true eigenmodes of the system. We suggest that these results complicate the conclusions arrived at by Lau & Binney (2021) on the importance of Case-van Kampen modes.

The theories of Landau damping and the Case-van Kampen modes were initially developed for a collisionless plasma with particle distribution functions described by the Vlasov equation. Landau (1946) demonstrated that plasma waves are damped by solving the initial-value problem for the DF using Laplace transforms on the Vlasov equation despite the absence of dissipation. However, these damped modes (or the Landau modes) are not true eigenmodes of the collisionless system. The complex frequencies for these modes do not satisfy a true dispersion relation for eigenmodes but are the long-time remnants of an initial perturbation. The frequencies of these modes are at the locations of complex poles, which appear when evaluating the inverse Laplace transform.

The basic set of eigenmodes referred to above as Case-van Kampen modes, were found by van Kampen (1955) and Case (1959) following Landau's theory. The set is a continuum with an eigenmode for every given real frequency ($\omega$) and real wavenumber ($k$). Individually, these modes are singular in form and purely oscillatory. Landau damping can be recovered by a superposition of a continuum of these modes resulting in a non-singular distribution, with the damping emerging as a consequence of the phase-mixing of the modes.

In reality, the collisionless assumption is simply the idealized limit of zero collisions. Thus, for a system characterized by a collision frequency, $\nu$, the collisionless case may be simply realized by setting $\nu = 0$. However, the collisional terms in the Boltzmann equation, as represented, for example, by the Fokker-Planck operator, involve second-order derivatives of the distribution function with respect to the particle velocity. Therefore, the weak collision limit $\nu \to 0$ is, *prima facie*, a singular perturbation. In other words, the collision terms cannot be ignored due to the development of large velocity gradients, and the limit $\nu \to 0$ can be drastically different from the collisionless case $\nu = 0$. Lenard & Bernstein (1958) (hereafter LB) considered the effect of weak collision for plasma waves using a simplified Fokker-Planck type collision operator. They showed that the complex frequency of the least damped eigenmode tends to that of the least damped Landau mode in the limit of zero collisions, but they did not discuss the effect of this limit on the underlying spectrum. This limit was considered by Ng, Bhattacharjee, & Skiff (hereafter NBS) (1999, 2004) by analysis and direct numerical calculations. NBS (1999) found that the eigenmode spectrum becomes discrete under the LB collision operator and that the Case-van Kampen spectrum ceases to exist. Furthermore, they showed that the discrete eigenmode spectrum includes the Landau modes and new discrete modes, absent in the collisionless theory. These findings were further confirmed analytically by NBS (2004), who also showed that the set of discrete eigenmodes is complete.

In Section 2, we will present the theory for linear waves in a weakly collisional stellar system. While the mathematical formulations for stellar and plasma systems are similar, there is a fundamental difference. Standard electrostatic plasmas, which obey Coulomb's law and are comprised of positive and negative charges, exhibit both attractive and repulsive forces and are quasi-neutral on the whole. In contrast, stellar systems obey Newton's law of gravitation which is always attractive, and an equilibrium state is necessarily spatially inhomogeneous. Relying on the so-called Jeans swindle (Binney & Tremaine 2008), we will consider the particular case of a stellar system that will be assumed to be spatially homogeneous in equilibrium. We will explicitly calculate eigenmode frequencies for both the collisionless and weakly collisional cases, including the discrete mode for the Jeans instability.

Applying the same formulation, we will discuss in Section 3 the great sensitivity of the eigenmode spectrum to small perturbations in the equilibrium distribution function in a weakly collisional system. This consideration is stimulated by the work of Skiff et al. (2002), who reported experimental and theoretical results showing that the spectrum of the eigenmodes of the waves can change drastically even by adding a small perturbation to a Maxwellian plasma. We will show by explicit calculations for a stellar system that a heavily damped system can become weakly damped even if a small shifted Maxwellian distribution is superimposed on the main Maxwellian DF.

We will conclude in Section 4 with a summary of our findings and possible implications of our results.

**2. Eigenmodes With and Without Weak Collision**

Let us start by considering a kinetic eigenmode of perturbations within a uniform background of a stellar system with weak collisions. The distribution function $F(\mathbf{x}, \mathbf{v}, t)$ satisfies the Boltzmann equation of the form

$$\frac{\partial F}{\partial t} + \mathbf{v} \cdot \frac{\partial F}{\partial \mathbf{x}} - \frac{\partial \Phi}{\partial \mathbf{x}} \cdot \frac{\partial F}{\partial \mathbf{v}} = \nu \frac{\partial}{\partial \mathbf{v}} \cdot \left( \mathbf{v} F + v_0^2 \frac{\partial F}{\partial \mathbf{v}} \right), \qquad (1)$$

where $\Phi$ is the gravitational potential. The LB collision operator on the right-hand side assumes that the collision frequency $\nu$ is constant and $v_0$ is the background thermal velocity. It is a special case of the Fokker-Planck operator, is more tractable analytically, and has been used in textbooks on stellar and galactic systems (cf. Saslaw 1985). The LB operator obeys the H-theorem and conserves particle number but does not conserve energy and momentum. The non-conservative properties could become physically important in the strong-collision limit. Since our focus here is on the weak-collision limit, our conclusions based on the LB operator should not be altered qualitatively by considering the full Fokker-Planck operator, which conserves particles, energy, and momentum.

We define the stellar mass density $\rho$ by integrating $F$ over velocity space, that is, $\rho = \int d^3v\, F$, which acts as a source term on the right-hand side of the self-consistent Poisson equation,

$$\nabla^2 \Phi = 4\pi G \int d^3v\, F, \tag{2}$$

where $G$ is the gravitational constant. For simplicity, we now consider a one-dimensional (1D) perturbation along the $x$-direction imposed on a Maxwellian background $F_0(\mathbf{v})$ so that the reduced perturbation $f(x, v, t)$ is defined by the relation

$$f(x, v, t) = \int dv_y dv_z (F - F_0) = \int dv_y dv_z F - f_0, \tag{3}$$

where we have dropped the subscript in $v_x$ and $f_0$ is the 1D Maxwellian distribution. Equations (1) and (2) can now be linearized to yield

$$\frac{\partial f}{\partial t} + v \frac{\partial f}{\partial x} - \frac{\partial \Phi}{\partial x} \frac{\partial f_0}{\partial v} = \nu \frac{\partial}{\partial v}\left(vf + v_0^2 \frac{\partial f}{\partial v}\right), \tag{4}$$

$$\frac{\partial^2 \Phi}{\partial x^2} = 4\pi G \int d\,vf. \tag{5}$$

For an eigenmode, we consider a perturbation of the form $f(x, v, t) = \tilde{f}(k, \omega, v) \exp[i(kx - \omega t)]$. We can consider two types of problems: (i) the temporal evolution problem, for which $k$ is real and $\omega$ is complex, and (ii) the spatial evolution problem, for which $\omega$ is real and $k$ is complex. Here we will consider (i). We define a set of dimensionless variables and combine Eqs. (4) and (5) to obtain

$$(u - \Omega)g(u) - \eta(u)\int_{-\infty}^{\infty} g(u')du' = -i\mu\frac{\partial}{\partial u}\left(ug + \frac{1}{2}\frac{\partial g}{\partial u}\right), \quad (6)$$

where $u = v/(\sqrt{2}v_0)$, $\Omega = \omega/(\sqrt{2}kv_0)$, $g = \sqrt{2}v_0 f/\rho_0$, where $\rho_0$ is the background mass density, $\eta(u) \equiv \alpha(\partial g_0/\partial u)/2$, with $\alpha = -4\pi G\rho_0/k^2 v_0^2$ and $g_0 \equiv \exp[-u^2]/\pi^{1/2}$, $\mu = \nu/(\sqrt{2}kv_0)$. (Note that the analogous electrostatic plasma wave problem can be described by the same Eq. (6) with a change in the definition $\alpha = \omega_p^2/(k^2 v_0^2)$, where $\omega_p$ is the plasma frequency.)

Before discussing the eigenmode solutions for Eq. (6) for the collisional problem $\mu \neq 0$, we briefly consider the collisionless case $\mu = 0$. It can be shown by direct substitution that for any real $\Omega$, there is a solution of the form

$$g_\Omega(u) = P\left[\frac{\eta(u)}{u - \Omega}\right] + \delta(u - \Omega)\left[1 - P\int_{-\infty}^{\infty}\frac{\eta(u')}{u' - \Omega}du'\right], \quad (7)$$

where $\delta$ denotes the Dirac delta function, and $P$ requires taking the principal part when performing integration along the real $u$ axis, which is the appropriate prescription for Case-van Kampen modes. In this case, the eigenvalues $\Omega$ constitute a continuous spectrum along the real axis. Equation (6) for $\mu = 0$ also admits discrete eigenvalues with complex $\Omega$. In this case, the solution is still given by Eq. (7) without the second term on the right-hand side and with the principal part dropped (since it becomes unnecessary) and $\Omega$ must satisfy the dispersion relation

$$1 - \int_{-\infty}^{\infty}\frac{\eta(u)}{u - \Omega}du = 0. \quad (8)$$

The eigenmodes for the continuum of real $\Omega$ plus possible discrete modes with complex $\Omega$ satisfying Eq. (8) were shown by van Kampen (1955) and Case (1959) to form a complete set of solutions. Individually, each mode described by Eq. (7) for a real $\Omega$ is singular in form and therefore cannot represent a physical perturbation in isolation. Instead, these modes have to be combined to yield a time-dependent solution

$$g(x, t, u) = e^{ix}\int_{-\infty}^{\infty} e^{-i\Omega t}c(\Omega)g_\Omega(u)d\Omega, \quad (9)$$

with $c(\Omega)$ determined by the initial condition

$$g(x, 0, u) = e^{ix}\int_{-\infty}^{\infty} c(\Omega)g_\Omega(u)d\Omega, \quad (10)$$

assuming that there is no discrete eigenmode. Moreover, since $\Omega$ is real, the perturbation described by Eq. (9) is undamped. To see the well-known effect of Landau damping, one can consider the effect of the $\delta$-function term in Eq. (7) when it is substituted into Eq. (9). After the integration in $\Omega$, we see that $g$ has a factor proportional to $\exp(-iut)$, which makes $g$ increasingly oscillatory in the $u$-space in the long-time limit. Therefore, the wave potential, proportional to $\int_{-\infty}^{\infty} g(u)\,du$ (obtained from the Poisson equation), will decay in time due to the cancellation of an increasingly oscillatory $g$ in the integrand (the Riemann-Lebesque Lemma). This effect is called phase-mixing. The Landau damping effect can be found by solving the initial-value problem using Laplace transforms such that

$$g(x,t,u) = e^{ix} \int_L e^{-i\Omega t} \tilde{g}(\Omega, u) \frac{d\Omega}{2\pi}, \tag{11}$$

with

$$\tilde{g}(\Omega, u) = \left[ \frac{\eta(u)}{D(\Omega)} \int_C \frac{g(x,0,u')}{u' - \Omega} du' + g(x,0,u) \right] / i(u - \Omega), \tag{12}$$

where

$$D(\Omega) = 1 - \int_C \frac{\eta(u)}{u - \Omega} du, \tag{13}$$

with the contour $C$ taken along the real $u$ axis if the imaginary part of $\Omega$ is positive, i.e., $\Omega_i > 0$, but deformed to encircle the pole $u = \Omega$ if $\Omega_i \leq 0$ (to ensure the proper analytical continuation of the function $D(\Omega)$ from the upper half $u$-plane to the lower half $u$-plane). Formally the integral in Eq. (11) can be expressed as

$$g(x,t,u) = e^{ix} \left[ G_0(u) e^{-iut} + \sum_{n=1}^{\infty} G_n(u) e^{-i\Omega_n t} \right] + \text{transient}, \tag{14}$$

where the $G_0(u)$ term is due to the pole at $\Omega = u$, and the $G_n(u)$ term is due to a pole at a Landau mode $\Omega = \Omega_n$, satisfying

$$D(\Omega_n) = 1 - \int_C \frac{\eta(u)}{u - \Omega_n} du = 0. \tag{15}$$

Equation (15) is the same as the dispersion relation Eq. (8) for $\Omega_{ni} > 0$, which corresponds to an unstable mode. However, Eq. (15) is not equivalent to Eq. (8) for $\Omega_{ni} \leq 0$ due to the deformation of the contour $C$ required by the Landau prescription. There is no solution to Eq. (8) for the plasma wave problem if the background DF is Maxwellian. Instead, there is an infinite set of discrete $\Omega_n$ with $\Omega_{ni} < 0$ that satisfies Eq. (15). All these Landau modes are damped, but these modes are not true eigenmodes because $\Omega_n$ do not satisfy the dispersion relation (8). The Landau mode equation (15) can also be written

$$D(\Omega_n) = 1 + \alpha[1 + \Omega_n Z(\Omega_n)] = 0, \tag{16}$$

for a Maxwellian DF, where $Z(\Omega)$ is the well-known plasma dispersion function. $Z(\Omega)$ can also be related to the complementary error function of complex argument by the identity $Z(\Omega) = i\sqrt{\pi} \exp(-\Omega^2) \operatorname{erfc}(-i\Omega)$.

In contrast with the plasma problem, for the stellar problem, we have $\alpha < 0$. Then it is possible to find a solution for Eq. (8) with $\Omega_i > 0$ for a Maxwellian DF. Since $\Omega_i > 0$, this true eigenmode is also a Landau mode satisfying Eq. (16) that happens to be a discrete Case-van Kampen mode as well. As it turns out, this mode is non-propagating, i.e., with the real part of its frequency $\Omega_r = 0$, which can be written alternately as $\Omega = i\Omega_i$. At the marginally stable point, the frequency of this mode is zero. From Eq. (16), the condition for $\Omega = 0$ is $\alpha = -1$. An expansion of the function $Z(\Omega)$ near $\Omega = 0$ in Eq. (16) gives

$$\Omega \approx \frac{i}{\sqrt{\pi}}\left(1 + \frac{1}{\alpha}\right) = \begin{cases} \text{stable} & \alpha > -1 \\ 0 & \text{for } \alpha = -1 \\ \text{unstable} & \alpha < -1 \end{cases}. \tag{17}$$

We infer that there is always an unstable mode if $\alpha < -1$. This mode is the well-known Jeans instability (Jeans 1902) (hereafter referred to as the Jeans mode). From the definition of the parameter $\alpha$, the condition $\alpha < -1$ can be rewritten $4\pi G \rho_0 > k^2 v_0^2$. From this inequality, one can easily see that the physical drive for instability is the gravitational attraction between the stars. For a Maxwellian DF, all Landau modes other than the Jeans mode have $\Omega_{ni} < 0$ and are damped (just as the plasma waves are). Figure 1 plots the locations of the frequencies $\Omega_n$ of the Landau modes solved from Eq. (16) for the case $\alpha = -10^{-4}$. In this case, by the criterion (17), we expect the Jeans mode to be damped. For ease of visualization, it is preferable to define $\kappa = 1/\Omega$ and plot the locations on the complex $\kappa$-plane. Note that $\kappa_r = \Omega_r/|\Omega|^2$ and $\kappa_i = -\Omega_i/|\Omega|^2$. Therefore, all modes shown in Fig. 1 are damped. It is also apparent that an infinite sequence of modes exists along the $\kappa_r = \kappa_i$ line in the $|\kappa| \to 0$ limit, i.e., the $|\Omega| \to \infty$ limit. All these modes are heavily damped. The mode that is on the $\kappa_r = 0$ axis is the Jeans mode, which is damped for this case.

While the stability condition for the Jeans mode given by Eq. (17) is derived from kinetic theory, the same condition can be obtained in a much simpler fluid treatment (Binney & Tremaine 2008), with $\Omega^2$ given by

$$\Omega^2 = (1+\alpha)/2 = \begin{cases} \text{stable} & \alpha > -1 \\ 0 & \text{for } \alpha = -1 \\ \text{unstable} & \alpha < -1 \end{cases}. \tag{18}$$

Note that the frequency predicted by fluid theory is quite different from Eq. (17) near $\alpha \approx -1$, although the stability condition is still predicted correctly by fluid theory. In Fig. 2, we plot $\Omega_i$ for the Jeans mode as a function of $\alpha$ from the kinetic theory by solving Eq. (16) numerically (solid curve), and the fluid result Eq. (18) (dashed curve) with the damping branch not shown for $\alpha < -1$. From the figure, as well as the large-$\Omega$ asymptotic behavior of Eq. (16), both the kinetic theory and fluid theory give $\Omega^2 \to \alpha/2$ for the unstable Jeans mode in the limit of $\alpha \to -\infty$.

Let us now consider the collisional problem, i.e., Eq. (6) with the normalized collision frequency $\mu \neq 0$. Following NBS (2004), the eigenfrequency $\Omega_n$ and eigenfunction $g_n$ for Eq. (6) can be solved by taking Fourier transforms in $u$-space, which gives

$$\left(w + \frac{1}{\mu}\right)\frac{\partial \tilde{g}_n(w)}{\partial w} + \left(\frac{w^2}{2} - \frac{i\Omega_n}{\mu}\right)\tilde{g}_n(w) - \frac{\alpha w \tilde{g}_n(0)}{2\mu} e^{-w^2/4} = 0, \tag{19}$$

where $\tilde{g}_n(w) = \int_{-\infty}^{\infty} g_n(u) e^{iwu} du/\sqrt{2\pi}$. Since Eq. (19) is a first-order differential equation in $w$, closed-form solutions can be found straightforwardly:

$$\tilde{g}_n(w) \equiv 1 - D(w, \Omega_n) = -\alpha\left[1 + \frac{i\Omega_n}{\mu} d\left(\frac{1}{2\mu^2} - \frac{i\Omega_n}{\mu}, \frac{1+\mu w}{2\mu^2}\right)\right] e^{-w^2/4}, \tag{20}$$

where $\tilde{g}_n(0) = 1$ is chosen, and so the eigenvalues satisfy dispersion relations

$$D_\mu(\Omega_n) \equiv D(0, \Omega_n) = 1 + \alpha\left[1 + \frac{i\Omega_n}{\mu} d\left(\frac{1}{2\mu^2} - \frac{i\Omega_n}{\mu}, \frac{1}{2\mu^2}\right)\right] = 0. \tag{21}$$

In these equations, the function $d$ is defined by $d(a,\zeta) \equiv \zeta^{-a} e^\zeta \gamma(a,\zeta)$, where $\gamma(a,\zeta) = \int_0^\zeta e^{-t} t^{a-1} dt$ is the incomplete gamma function, so that it is a single-valued analytic function in the complex $a$ and $\zeta$ planes, except for simple poles when $a$ is a non-positive integer, or $\Omega_{\mathrm{LB}n} = -i[n\mu + 1/(2\mu)]$ for a non-negative integer $n$ (the LB poles). From Eq. (7), we see that if $\Omega_n$ is an eigenvalue, so is the negative of its complex conjugate, $-\Omega_n^*$, with corresponding eigenfunction $g_n^*(-u)$. Note also that $\mathrm{Im}(\Omega_n) < 0$ for all $n$, i.e., all the modes are damped except the Jeans mode for $\alpha < -1$.

NBS (2004) demonstrate that the eigenfrequencies solved from Eq. (21) with corresponding eigenfunctions given by Eq. (20) form a complete set. (The proof is given for the plasma wave problem but should also hold in the present case with minor changes.) Here, we will limit our discussion to some properties of the set of eigenfrequencies, especially in the limit of weak collision $\mu \to 0$. Following an asymptotic expansion for the incomplete gamma function (NBS 2004), as well as the well-known asymptotic expansion (the Stirling's formula) for the gamma function, the $d$ function in Eq. (21), with $a = 1/2\mu^2 - i\Omega/\mu$ and $\zeta = 1/2\mu^2$, in the $\mu \to 0$ limit becomes

$$d(a,\zeta) \to \sqrt{\frac{\pi}{2a}} \exp\left[\frac{(a-\zeta)^2}{2a}\right] \mathrm{erfc}\left(\frac{a-\zeta}{\sqrt{2a}}\right), \qquad (22)$$

which is valid in the limit of $|a| \sim |\zeta| \to \infty$, but with a finite $(a-\zeta)/\sqrt{2a}$ and with $a$ not equal to a negative integer or zero. Therefore, the dispersion relation in the $\mu \to 0$ limit becomes

$$D_\mu(\Omega) \to 1 + \alpha\left[1 + i\Omega\sqrt{\pi}\exp(-\Omega^2)\mathrm{erfc}(-i\Omega)\right] = 1 + \alpha[1 + \Omega Z(\Omega)] = D(\Omega) = 0, \quad (23)$$

where the condition of $a$ not equal to a negative integer or zero means that $\Omega$ is not a LB pole. Thus, a part of the spectrum of eigenfrequencies for the collisional problem actually tends to the set of frequencies of the Landau modes in the weak-collision limit. The Landau modes, which are not eigenmodes (except for the unstable Jeans mode) for the collisionless problem ($\mu = 0$), now become true eigenmodes for the collisional problem in the limit $\mu \to 0$. Moreover, the Case-van Kampen modes, which are originally eigenmodes for the collisionless problem, now appear completely erased. This drastic change in the characteristics of the spectrum of the eigenfrequencies is a consequence of the fact that the weak-collision limit is a singular perturbation.

It is important to recognize that another part of the sequence of eigenfrequencies tends to the LB poles in the weak-collision limit so that there is no contradiction to the derivation of Eq. (23). To see why this is so, we may express the collisional dispersion relation formally as

$$D_\mu(\Omega) \xrightarrow{\mu \to 0} D(\Omega) + \mu D_1(\Omega) , \tag{24}$$

since we have established that $D_\mu \to D$ when $\mu \to 0$. Since $D_\mu$ is singular at the LB poles, but $D$ is regular, $D_1$ must be singular. Therefore, to satisfy $D_\mu = 0$ but $D \neq 0$, $\Omega$ must be near a LB pole, so that $D_1$ is large and $D + \mu D_1 = 0$ for $\mu \to 0$. From the functional forms of the LB poles, $\Omega = -i[n\mu + 1/(2\mu)]$ for $n$ being a positive integer or zero, all these modes are heavily damped.

To illustrate the analytical result shown in Eq. (23), we will focus on the eigenfrequency of the Jeans mode for the collisional problem. While it is possible to calculate the frequency from the dispersion relation, Eq. (21), in practice we calculate it using a numerical method based on Hermite polynomial expansion as presented by NBS (1999). In Fig. 3, we plot $\Omega_i$ for the Jeans mode as a function of $\alpha$ from the collisional kinetic theory (solid curve) for a small collision frequency of $\mu = 0.05$. This mode is still non-propagating with $\Omega = i\Omega_i$. By adding a simple Krook collision term, the fluid theory can be modified to give the frequency of the Jeans mode as

$$\Omega = \frac{1}{2}\left[-i\mu \pm \sqrt{-\mu^2 + 2(1+\alpha)}\right] . \tag{25}$$

From Eq. (25), it is clear that the instability condition is still $\alpha < -1$, as in the collisionless case. The same instability condition holds for the kinetic case, as shown in Fig. 3. The $\Omega_i$ for the Jeans mode from the fluid theory is also plotted for the unstable branch (dashed curve) and stable branch (dashed-dotted curve). The numerical difference between Fig. 2 and Fig. 3 is very small. One difference is that the fluid mode for $\alpha > -1$ now has small damping with $\Omega_i = -\mu/2$. However, the nature of the mode is very different. The collisional mode in Fig. 3 shown as the solid curve is a true eigenmode for all $\alpha$. The corresponding curve in Fig. 2 is an eigenmode for the collisionless problem only for $\alpha < -1$ when the mode is unstable. For $\alpha > -1$, it is just the damping rate for the Landau mode, which is not an eigenmode. We can look at this situation more closely in Fig. 4 by plotting $\Omega_i$ as a function of $\mu$ for $\alpha = -0.99$ (solid curve). The dashed curve in Fig. 4 is one of the two damped modes from Eq. (25) based on the fluid theory. The value of $\Omega_i$ for the solid curve at $\mu = 0$ is the same as calculated from the collisionless theory, Eq. (16). While it may seem difficult to confirm this by reading off the value from Fig. 2, the data underlying the figure do show the agreement. Figure 4 also shows that $\Omega_i$ both the kinetic and the fluid calculations agree with each other, including that its magnitude tends to zero in the large $\mu$ limit. This behavior can be shown from Eq. (25) in the strong-collision limit with $\mu^2 \gg |1 + \alpha|$,

$$\Omega \xrightarrow{\mu \to \infty} \begin{cases} -i(1+\alpha)/2\mu \\ -i\mu \end{cases} = \begin{cases} \text{both stable} & \alpha > -1 \\ \text{one zero, one stable} & \text{for} \quad \alpha = -1 \\ \text{one stable, one unstable} & \alpha < -1 \end{cases}. \tag{26}$$

Figure 5 shows a plot similar to Fig. 4 but for the unstable case with $\alpha = -1.01$. Since this mode is unstable, it is an eigenmode for both the collisionless problem and the collisional problem, with the $\Omega_i$ value tending to the collisionless value in the $\mu \to 0$ limit. The strong-collision limit for both kinetic and fluid calculations is again given by Eq. (26).

## 3. Non-Maxwellian Background

Our discussion so far has been confined to the assumption of a Maxwellian background DF. However, in a weakly collisional system, the background can easily deviate from a Maxwellian DF. We have shown in the last section that even a very weak collision can have physical consequences that are drastically different from those under the idealized collisionless assumption. Similarly, a small deviation from the Maxwellian background can produce large changes in physical consequences. We will demonstrate this effect by considering a non-Maxwellian case with the background DF $g_0$ of the following form:

$$g_0 = \frac{1}{(1+\varepsilon\xi^2)\sqrt{\pi}}\{\exp(-u^2) + \varepsilon\xi^3 \exp[-\xi^2(u-\bar{u})^2]\}, \tag{27}$$

which represents a shifted tail population superimposed on the main Maxwellian DF. Note that $g_0$ is normalized such that $\int g_0 du = 1$. Since we have established in the last section that the Landau modes become true eigenmodes in the weak-collision limit, we will calculate the eigenfrequencies using the collisionless treatment. Therefore, the dispersion relation is simply generalized from Eq. (16) to, using the plasma dispersion function $Z$,

$$1 + \frac{\alpha}{(1+\varepsilon\xi^2)}\left(1 + \Omega Z(\Omega) + \varepsilon\xi^4\{1 + \xi(\Omega - \bar{u})Z[\xi(\Omega - \bar{u})]\}\right) = 0. \tag{28}$$

Figure 6 shows the locations of eigenfrequencies solved from Eq. (28) in the complex $\kappa = 1/\Omega$ plane, for the case with $\alpha = -10^{-4}$, $\varepsilon = 0.0001$, $\xi = 5$, $\bar{u} = 7$. We see that by adding a tiny tail component, the spectrum of eigenfrequencies becomes very different from that in Fig. 1. Mainly, there is a totally new branch in addition to the old one. On this new branch, there are modes with small damping near $\kappa_r \approx 0.14$. This new branch develops due to the small perturbation in the far tail of the background DF. The population of particles in this perturbation term is only 0.0025 of the main DF. We can find similar results if we perturb the main body of the background distribution. Figure 7 shows the eigenfrequencies for the case with $\alpha = -0.2$, $\varepsilon = 0.0005$, $\xi = 5$, $\bar{u} = 1.5$. We see that there are still weakly damped modes, especially around $\kappa_r \approx 0.6$, which are not there for the Maxwellian case with $\varepsilon = 0$. Figure 8 shows $g_0$ for this case, which deviates from a Maxwellian (dotted curve) only slightly near $u = \bar{u} = 1.5$ with population about 0.0125 of the main distribution. The generation of these weakly damped modes cannot be explained purely by the nonzero average velocity of the background distribution $u_0 \equiv \int u\, g_0 du = \varepsilon\xi^2\bar{u}/(1+\varepsilon\xi^2)$ since this value is still very small for the above two cases: 0.0175 for Fig. 6, and 0.0185 for Fig. 7. While we are only showing these two examples with all modes remain damped, unstable modes can be excited, of course, with an even more significant deviation from the Maxwellian DF. The very existence of weakly damped modes can have critical physical consequences on the statistics of thermal fluctuations, compared to the case in which all modes are heavily damped.

We can attempt to describe the above results by a two-fluid theory, with one fluid corresponding to the main DF and the other corresponding to a weak "beam-like" component. The equations of motion and continuity equation for the two species $s$ are

$$\rho_s\left(\frac{\partial v_s}{\partial t} + v_s \frac{\partial v_s}{\partial x}\right) = -\rho_s \frac{\partial \Phi_1}{\partial x} - \frac{\partial p_s}{\partial x}, \tag{29}$$

$$\frac{\partial \rho_s}{\partial t} + \frac{\partial (\rho_s v_s)}{\partial x} = 0, \tag{30}$$

where $s = m, b$ represent the main and the beam-like components, $\rho_s$, $v_s$, and $p_s$ is mass density, flow velocity, and pressure respectively. We then linearize Eqs. (29) and (30) by introducing zeroth and first order quantities $\rho_s = \rho_{s0} + \rho_{s1}$, $v_m = v_{m1}$, $v_b = v_{b0} + v_{b1}$. Then, $\partial^2 \Phi_1 / \partial^2 x = 4\pi G(\rho_{m1} + \rho_{b1})$ and $\partial p_s / \partial x = v_{Ts}^2 \partial \rho_{s1} / \partial x$ where $v_{Ts}^2$ is the square of the thermal velocity of the species $s$. Taking the first order quantities to be proportional to $\exp[i(kx - \omega t)]$ we obtain

$$\rho_{m1} = \frac{k^2 \rho_{m0} \Phi_1}{\omega^2 - k^2 v_{Tm}^2}, \tag{31}$$

$$\rho_{b1} = \frac{k^2 \rho_{b0} \Phi_1}{(\omega - k v_{b0})^2 - k^2 v_{Tb}^2}. \tag{32}$$

Define the total density $\rho_0$ such that $\rho_{m1} = r\rho_0$, $\rho_{b1} = (1-r)\rho_0$. Note that $r = 1/(1 + \varepsilon \xi^2)$, based on the notation used for Eq. (27). We can then solve for the dispersion relation by substituting Eqs. (31) and (32) into the Poisson equation $k^2 \Phi_1 = -4\pi G(\rho_{m1} + \rho_{b1})$. We end up with a fourth-order algebraic equation for $\kappa$,

$$a\kappa^4 + b\kappa^3 + c\kappa^2 + d\kappa + 1 = 0, \tag{33}$$

with

$$a = \frac{1}{2}\left(\frac{1}{2\xi^2} - \bar{u}^2\right)(r\alpha + 1) + (1 - r)\frac{\alpha}{4},$$
$$b = \bar{u}(1 + r\alpha), \quad c = \bar{u}^2 - \frac{1}{2}\left(\frac{1}{\xi^2} + 1 + \alpha\right), \quad d = -2\bar{u}, \tag{34}$$

where we define $\xi = v_{Tb}/v_{Tm}$, $\bar{u} = v_{b0}/\sqrt{2}kv_{Tm}$, $\kappa = 1/\Omega = \sqrt{2}kv_{Tm}/\omega$, and $\alpha = -4\pi G\rho_0/k^2 v_{Tm}^2$. Eq. (33) can be solved numerically. For the case in Fig. 6, the four roots of $\kappa$ are found to be, 1.414, 1.414, 0.140, and 0.146. The last two roots are close to the real part of the least damped roots in Fig. 6. For the case in Fig. 7, the four roots are found to be -1.579, 1.578, 0.611, 0.734. The last two roots are also close to the least damped roots in Fig. 7. As discussed in the last section, there are some other points of agreement between the fluid results and the Vlasov results, including the onset of instability. However, the problem with the fluid theory is that there is no criterion to determine which roots obtained from the fluid theory correspond to a true Landau mode. Also, the fluid theory does not yield any information on the damping of an oscillatory mode, which is essentially a kinetic effect.

## 4. Conclusions

In this paper, we have addressed the question of eigenmodes that constitute thermal fluctuations in a stellar system: are they Case-van Kampen or Landau? We have demonstrated in Section 2, using the Lenard-Bernstein collision operator, that while the Case-van Kampen modes are the eigenmodes in a collisionless system, the Landau modes are the eigenmodes of the system in the limit of zero collisions. Mathematically, this is a consequence of singular perturbation theory since the small parameter, the collision frequency, multiplies the highest-order velocity derivative of the kinetic equation. Thus, the zero-frequency limit of the collisional problem is drastically different from the problem without collisions. If this is a surprise, one should be prepared for it in the same way as one is if one takes the zero-viscosity limit of the Navier-Stokes equation or the zero-resistivity limit of the resistive magnetohydrodynamic equations.

While our demonstration uses the Lenard-Bernstein collision operator, the elimination of the Case-van Kampen modes as eigenmodes should happen for more general forms of collision operators that depend on derivatives of the distribution function. On the other hand, if discrete eigenmodes exist for a more general collision operator, part of the spectrum of eigenfrequencies should tend to the frequencies of Landau modes. An explicit demonstration of the results in this paper would be technically more difficult for a more general operator such as the Balescu-Lenard operator (Heyvaerts 2010, Chavanis 2012).

We hasten to caution the reader that one should not jump to the conclusion that the Case-van Kampen eigenmodes are not valuable for a statistical theory. Whether they are or not is a much subtler question beyond the scope of the present paper. From our perspective, it seems that a collisionless system may be thought of as either one that obeys the Vlasov equation or the zero-collision limit of a collision operator. Which of these perspectives holds may depend on the time scales of dynamical evolution. We do know that the mathematical structure of the collisional equations is profoundly different from those of the collisionless equations, which are well-known to be Hamiltonian. This raises the question of how many of the denumerably infinite constraints of the Vlasov equation survive in the weakly collisional limit. For example, the Boltzmann entropy is conserved by the Vlasov equation but not by the Lenard-Bernstein operator, which obeys the H-theorem.

While a small level of collision completely changes the nature of the spectrum of eigenmodes, the very weak level of collision may also have the effect of causing the system to deviate from a Maxwellian DF. With the Landau modes identified as true eigenmodes in a weakly collisional system, we have also considered the changes in the spectrum of the Landau eigenmode frequencies for a background slightly different from a Maxwellian DF in Section 3. The Landau modes for a Maxwellian DF are quite heavily damped, except for the possibility of the Jeans mode, which can become unstable for a specific range of parameters. We have shown by explicit calculations that the spectrum can change substantially if a tiny shifted Maxwellian is added to an existing Maxwellian background. Modes that are much weakly damped can appear then. Other cases not shown in this paper can have modes driven unstable with even more significant perturbations, even though the Jeans mode is stable for that parameter regime. However, even weakly damped modes can be important in a statistical theory since they dominate the spectrum of fluctuations.

**Acknowledgements**

This research is supported by a National Science Foundation grant PHY-2010617 and the Department of Energy award DE-AC02-09CH11466.

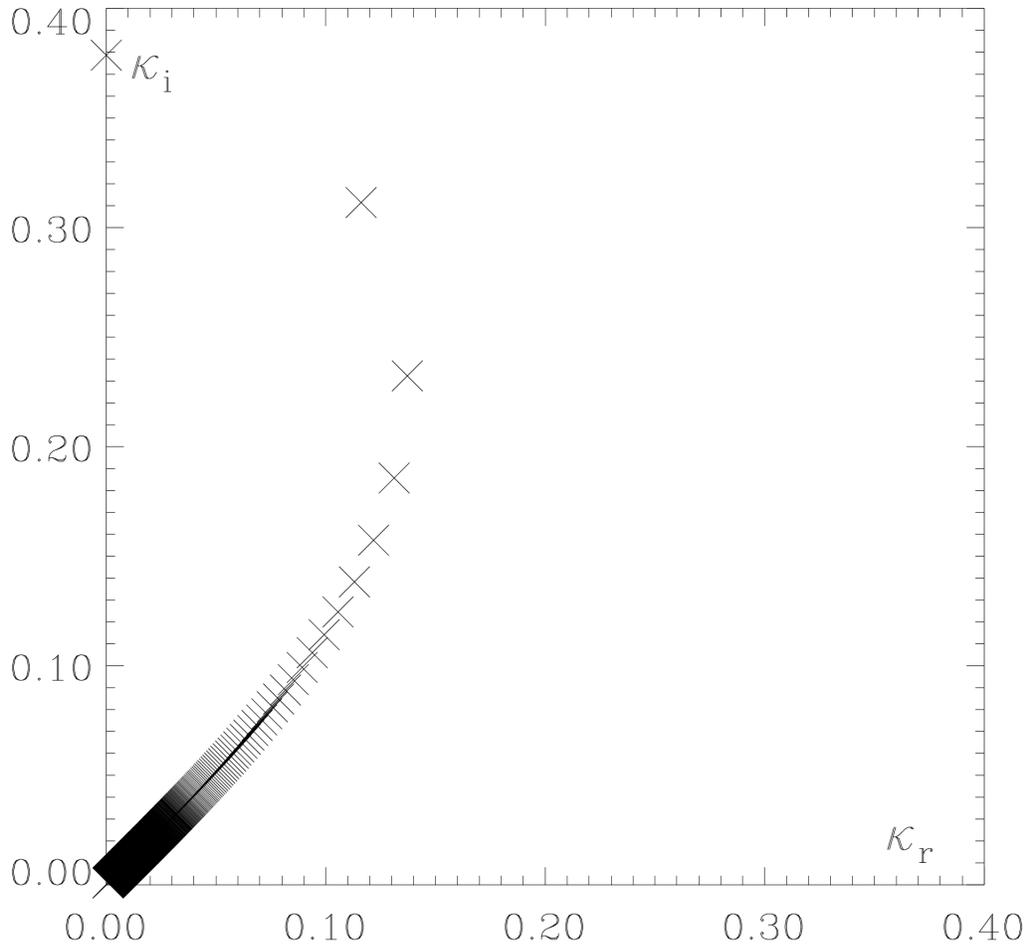

Figure 1. Locations (crosses) of the frequencies of the Landau modes for the $\alpha = -10^{-4}$ case on the complex plane of $\kappa = 1/\Omega$.

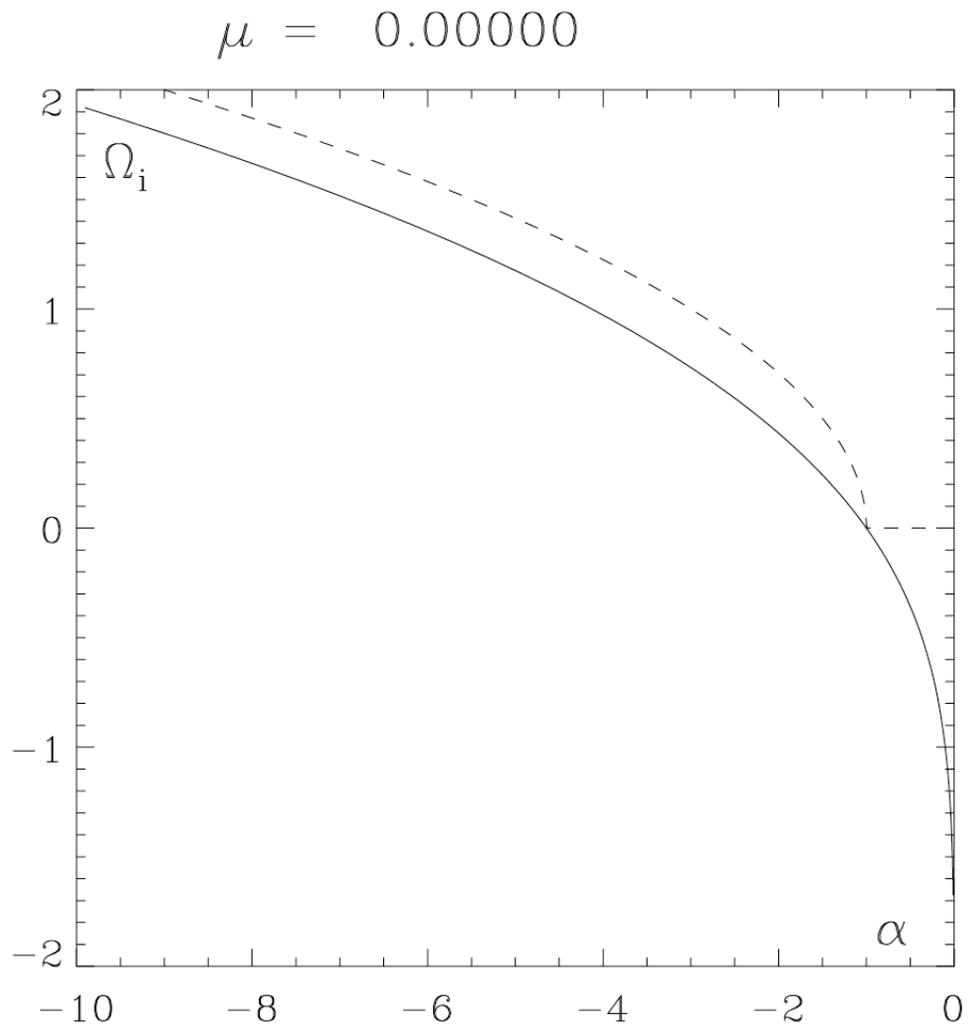

Figure 2. $\Omega_i$ of the Jeans mode as a function of $\alpha$ from the kinetic theory (solid curve) and the fluid theory (dashed curve) for the collisionless case with $\mu = 0$.

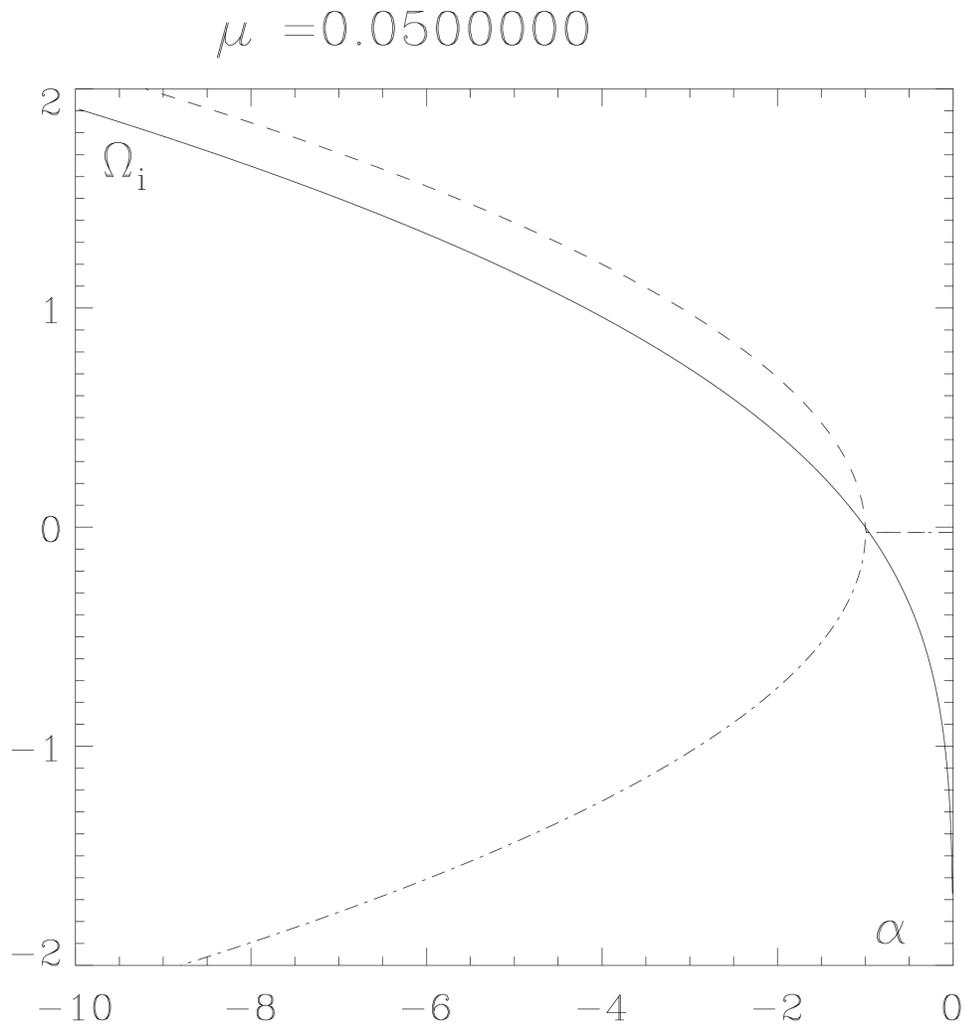

Figure 3. $\Omega_i$ of the Jeans mode as a function of $\alpha$ from the kinetic theory (solid curve), and the unstable branch from the fluid theory (dashed curve) and the stable branch (dashed-dotted curve), for the collisional case with $\mu = 0.05$.

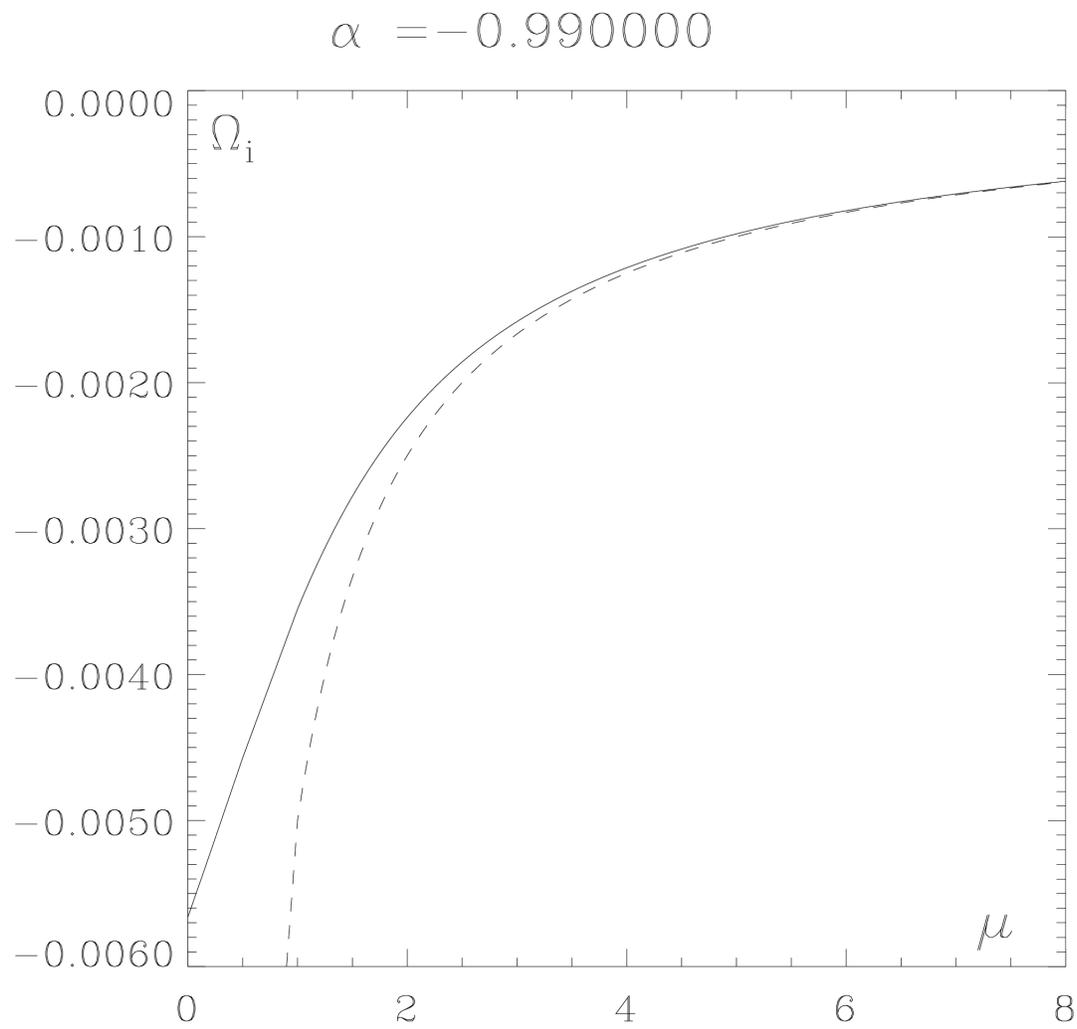

Figure 4. $\Omega_i$ of the Jeans mode as a function of $\mu$ for $\alpha = -0.99$ from the collisional kinetic theory (solid curve), and from the fluid theory (dashed curve).

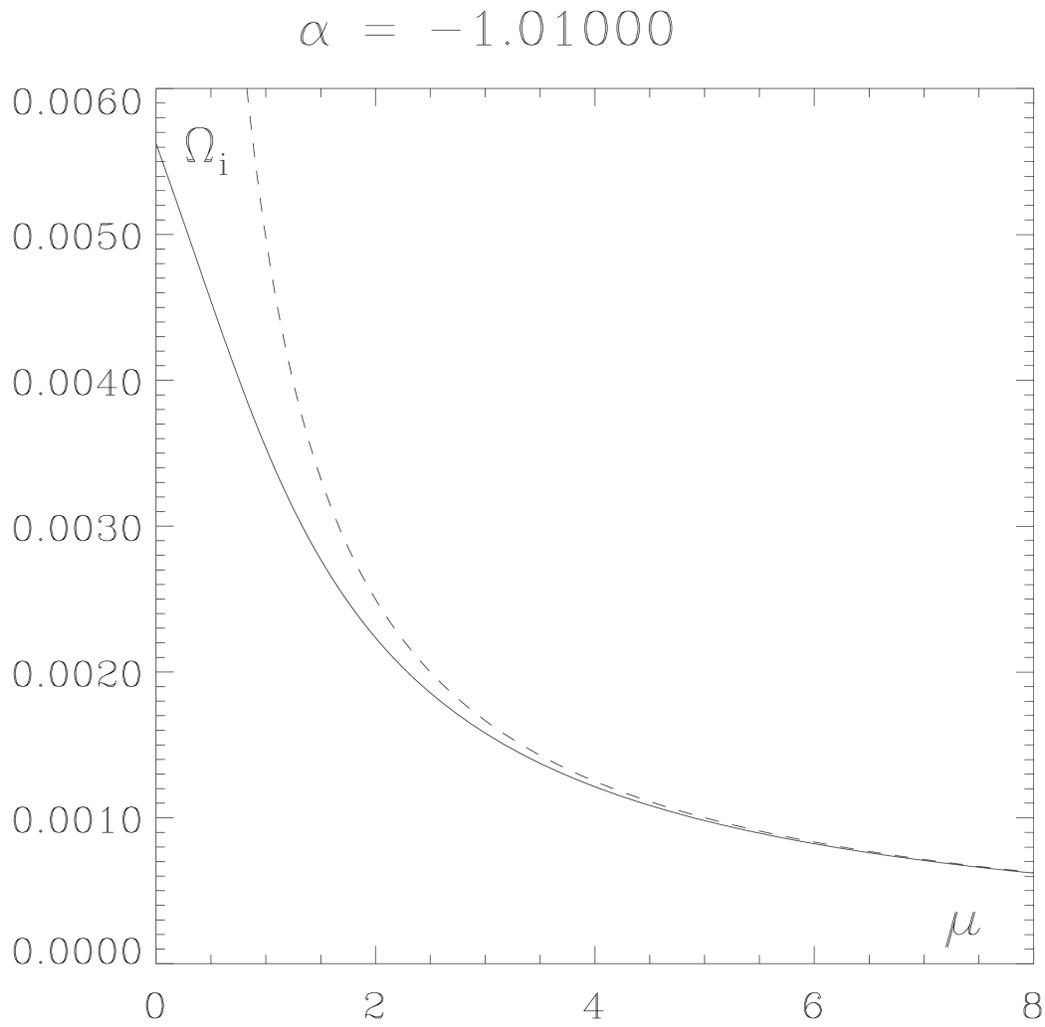

Figure 5. $\Omega_i$ of the Jeans mode as a function of $\mu$ for $\alpha = -1.01$ from the collisional kinetic theory (solid curve), and from the fluid theory (dashed curve).

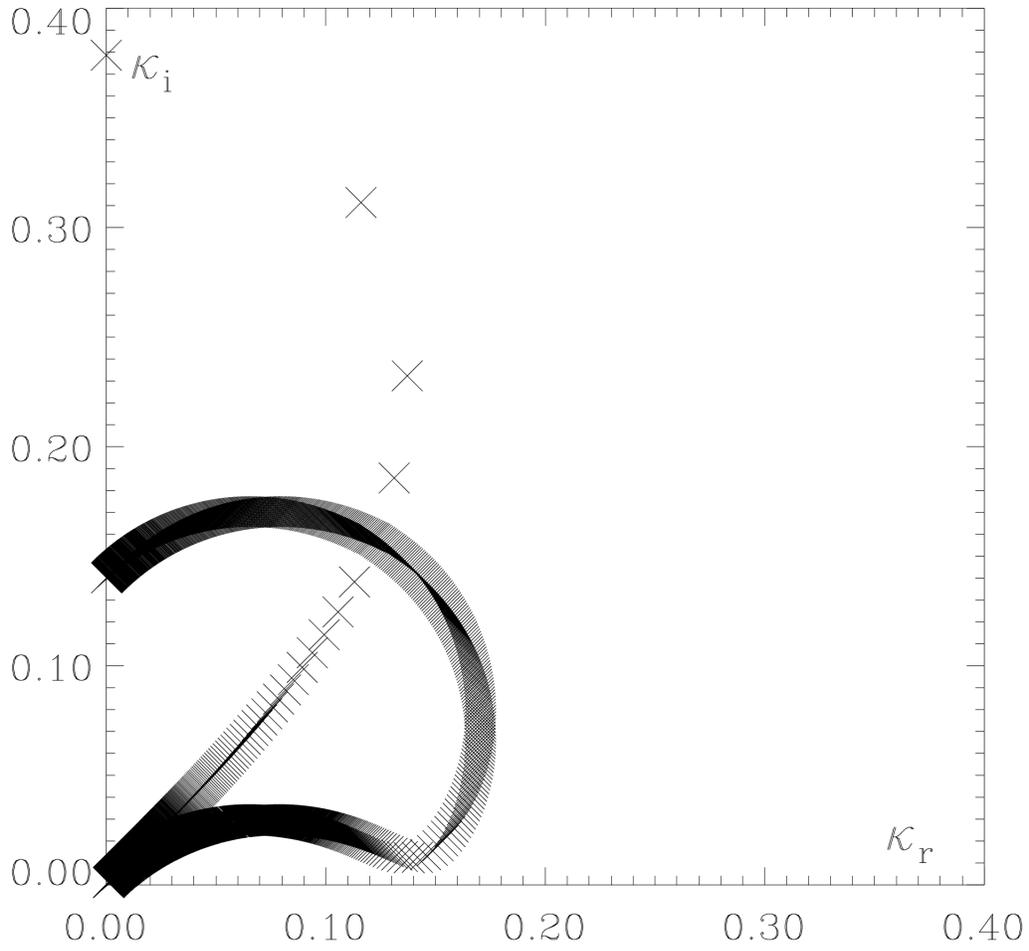

Figure 6. Locations of the frequencies of the Landau modes for the case with $\alpha = -10^{-4}$, $\varepsilon = 0.0001$, $\xi = 5$, $\bar{u} = 7$.

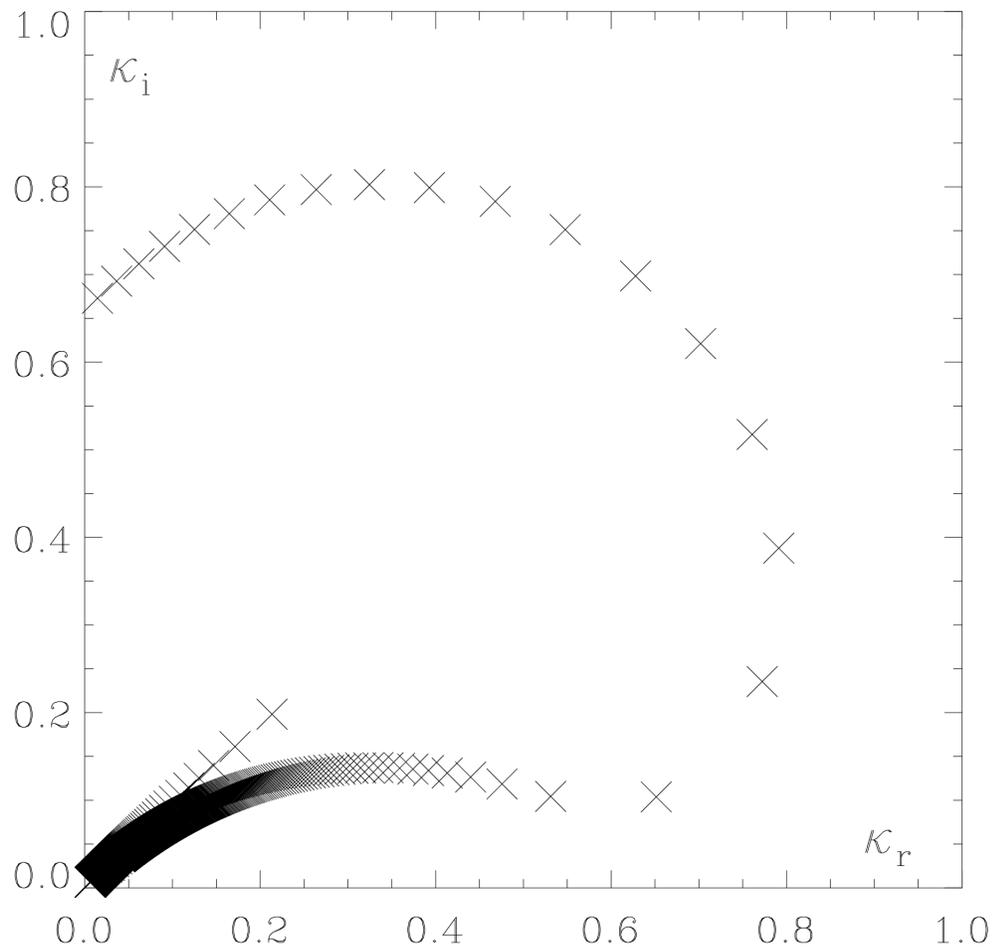

Figure 7. Locations of the frequencies of the Landau modes for the case with $\alpha = -0.2$, $\varepsilon = 0.0005$, $\xi = 5$, $\bar{u} = 1.5$.

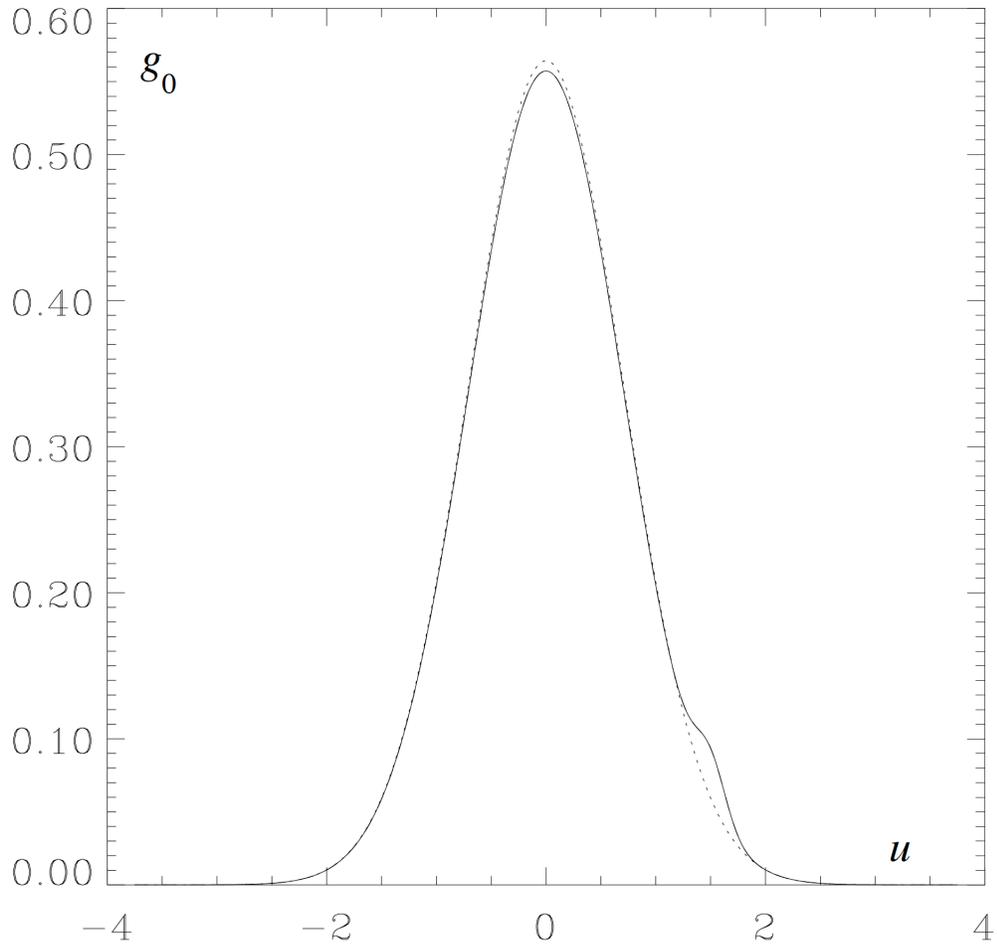

Figure 8. The background distribution function $g_0$ as a function of velocity $u$ for the case shown in Fig. 7. The main Maxwellian component of $g_0$ is shown as the dotted curve after normalization.